\documentclass[11pt]{article} 
\usepackage{graphicx} 
\usepackage{epsfig} 
 
\def\laq{~\raise 0.4ex\hbox{$<$}\kern -0.8em\lower 0.62ex 
\hbox{$\sim$}~} \def\gaq{~\raise 0.4ex 
\hbox{$>$}\kern -0.7em\lower 0.62ex 
\hbox{$\sim$}~}

\newcommand{\beq}{\begin{equation}} 
\newcommand{\eeq}{\end{equation}} 
\newcommand{\bea}{\begin{eqnarray}} 
\newcommand{\eea}{\end{eqnarray}}

\begin{document} \begin{titlepage} 
\begin{flushright} 
CERN-PH-TH/2004-065\\ 
ROM2F/2004/11\\
DESY 04-068\\ 
hep-th/0404262 
\end{flushright} 
\vspace*{1cm} 
\begin{center} 
\Large{\bf  Isospin Mixing of Narrow Pentaquark States} 
\end{center} 
\vspace*{1cm} 
 
\centerline{ {\large{G.C. Rossi}}\,$^{{a)}}$ 
and {\large{G. Veneziano}}\,$^{{b)}}$} 
\vskip 0.5cm 
\centerline{$^{a)}${\small{Dipartimento 
di Fisica, Universit\`a di  Roma ``{\it Tor Vergata}'', Italy}}} 
\centerline{\small{INFN, Sezione di Roma ``{\it Tor Vergata}'', Italy}} 
\centerline{\small{and}} 
\centerline{{\small{John von Neumann Institute for Computing, DESY Zeuthen, 
Germany}}} 
\smallskip 
\centerline{${}^{b)}${\small{Department of Physics, Theory Division}}} 
\centerline{\small{CERN, CH-1211 Geneva 23, Switzerland}} 

\vspace*{5mm} 
\begin{center} 
\begin{abstract} 
Interpreting the recently discovered narrow exotic baryons as pentaquark 
states, we discuss, along an old argument of ours, the isospin mixing 
occurring within the two doublets of $Q = -1$ and $Q=0$  states  lying 
inside the $S=-2$ ($\Xi$-cascade) sector. We argue that, at least within 
the Jaffe-Wilczek assignment, presently available data already indicate 
that mixing should occur at an observable level in both charge sectors, 
with  mixing angles that can be predicted in terms of ratios of observable 
mass splittings. 
\end{abstract} 
\end{center} 
\end{titlepage} 
 
\section{Introduction} 
\label{Sec1} 
 
Thirty five years after their existence had been predicted from duality 
arguments~\cite{Rosner}, claims that exotic hadrons have been discovered 
in various channels as narrow baryonic states, have been made by a number 
of experiments~\cite{EXP}. Although care must be exerted in their 
interpretation~\cite{DKS,FW}, evidence in favour of the existence of 
several such  states has been steadily mounting in recent months and will 
be accepted hereafter. 
 
The old, pre-QCD argument of~\cite{Rosner} was reconsidered, some ten 
years later, on the basis of a simple topology-based expansion of 
QCD~\cite{RV1,MRV}. It was argued that, from that more modern viewpoint, 
duality in $B\bar B$ scattering has to be interpreted as a relation between 
states in scattering and annihilation channels. More precisely, QCD baryons 
resemble Y-shaped strings with quarks  at the three ends and a junction, 
keeping track of baryon number, sitting in the middle. To leading order 
in the expansion, annihilation channels in one-, two-, or three-$q\bar q$ 
resonances are dual to channels where three new families of mesons 
(collectively called ``baryonia") are exchanged. They correspond, 
respectively, to bosonic states with four, two, and no quarks, and two 
string junctions. Baryonia decay preferentially (i.e.\ through string 
breaking) in $B\bar B$ channels (thus conserving the number of junctions), 
hence were argued to be narrow if near or below the $B\bar B$ threshold. 
 
Generalization of standard duality arguments allowed to make predictions 
for baryonium intercepts and slopes and led to an elegant explanation for 
the seemingly large breaking of exchange degeneracy of baryon trajectories. 
In the scheme there is naturally (or even necessarily~\cite{RV1,MRV}) 
room for baryonic pentaquark states of the type found in the experiments 
mentioned above. Having a dynamically favoured decay into a $B\bar B B$ 
final state, sufficiently light pentaquark states are also naturally narrow. 
The reason why preliminary evidence for narrow baryonia~\cite{MRV} was never 
confirmed is not completely clear to us. If accepted as an experimental fact, 
the non existence of  such states  would mean that baryonia, that are 
sufficiently heavy for their decay into mesons to be suppressed, are already 
significantly above the $B\bar B$ threshold. The situation would be 
different for the pentaquarks because of the higher threshold for decay 
in their preferred (three baryon) channels. 
 
Last but not least, mixings among nearby tetraquark  baryonium  states, 
resulting in large violations of isospin symmetry, were shown to be possible 
for sufficiently narrow states~\cite{RV2}. This phenomenon was argued to be 
quite general: indeed, at the very end of~\cite{RV2}, it was suggested 
that appreciable isospin mixing should occur in other exotic multiquark 
systems, such as pentaquarks. 
 
Multiquark states of the pentaquark type were also predicted in 
the chiral soliton model~\cite{CHIRSOL}, together with many other 
exotic baryonic states with increasingly large spin and isospin 
quantum numbers. 
 
Naturally the discovery of genuinely exotic resonances has 
prompted a good deal of new interesting investigations on possible 
dynamical mechanisms that would provide a theoretical basis for 
the existence and the narrowness of families of pentaquark states 
in QCD. Models of different kinds have been proposed complementing 
or in alternative to the old ideas of refs.~\cite{RV1} 
and~\cite{CHIRSOL}. Among them we should quote the diquark model 
of ref.~\cite{JW} to which we will come back in the next section 
and the works of refs.~\cite{KL} and~\cite{BS}. In~\cite{KL} the 
narrowness of the $\Theta^+$ states is explained as an 
interference effect between two almost degenerate states, while 
in~\cite{BS} it is suggested that the flavour structure of the 
$\Theta^+$ wave-function is such that after the meson is formed 
the residual three-quark piece has little overlap with the octet 
baryon wave-function (actually orthogonality becomes exact in the 
flavour $SU(3)$ symmetry limit). 
 
In this note we extend the idea of mixing among narrow states 
suggested in ref.~\cite{RV2} for tetraquarks to the case of 
the so-called $\Xi$-cascade, taking for granted the classification 
of these states suggested in ref.~\cite{JW}. It should be stressed, 
however, that the arguments we shall develop are rather general and 
apply even if the detailed pattern of degeneracies of the $\Xi$ states is 
different from the one advocated in~\cite{JW}. The key condition 
which is required for our analysis to work is the existence of 
almost degenerate states that can mix with a mass matrix, ${\cal M}$,  
in which electromagnetic and $(m_d - m_u)$ effects are comparable 
to (if not larger than) those due violations of the so-called OZI (Zweig) 
rule~\cite{OZIR}. Details of the predictions, however, will depend on the 
classification scheme one adopts and can be used, in principle, to 
distinguish and confront different dynamical models.

With inputs taken from experiments, and reasonable guesses about 
other, presently unknown, parameters entering the mass matrix, 
we find that, in the $Q=-1$ as well as in the $Q=0$ charge 
sector, the eigenstates of ${\cal M}$ should be appreciably different 
from pure isospin states, while an  almost maximal violation of isospin 
symmetry ($SU(2)$ ``ideal" mixing) is all but excluded. Such a theoretical 
picture has obvious experimental consequences for the production 
mechanisms and the decay of these narrow states. 
 
\section{A classification of $\Xi$ pentaquark states} 
\label{Sec2} 
 
The recently discovered $\Xi$ states can be classified, within the 
scheme proposed in ref.~\cite{JW}, as belonging to an ideally 
mixed ${\bf 8} + {\bf \bar{10}}$ of flavour $SU(3)$. In this scheme the $\Xi$ 
states consist of five valence quarks arranged as $(qs)(qs)\bar{q}$ 
where $q$ stands for $u$ or $d$, and the two-diquark system belongs to an 
isospin triplet (i.e.\ flavour-symmetric) state. As a result one has a total 
of six states with charge $Q$ going from $-2$ to $+1$. Out of these the 
sectors with $Q=-2$ and $Q=+1$ are pure $I=3/2$ states, while 
those with $Q=-1$ and $Q=0$ appear in pairs with  both $I=1/2$ 
and  $I=3/2$ components. The states with a given quark content are 
those with a definite $I_3$ of the two diquark system and of the 
antiquark. We shall use the following notation: 
\bea &&\Xi_d^{-} = |-1> | +1/2> \,\sim (ds)(ds)\bar{d} \, ,\qquad 
\Xi_u^{-} =|0>|-1/2>\, \sim (ds)(us)\bar{u} \;  , \nonumber  \\ 
&&\Xi_u^{0} =  |+1> | -1/2> \,\sim (us)(us)\bar{u} \, ,\qquad 
\Xi_d^{0} = |0> |+1/2>\, \sim (us)(ds)\bar{d} \, . 
\label{quarkstates} \eea The pure isospin states are related to 
those of~(\ref{quarkstates}) by standard Clebsh-Gordan (CG) 
coefficients. Using the conventions of the PDG one finds: 
\bea &&\Xi_d^{-} = \frac{1}{\sqrt{3}} \,[\Xi_{3/2}^{-}  - \sqrt{2} 
\Xi_{1/2}^{-} ] \, , \quad \Xi_u^{-} = \frac{1}{\sqrt{3}} 
\,[\Xi_{1/2}^{-}   + \sqrt{2} \Xi_{3/2}^{-} ] \, , \nonumber \\&& 
\Xi_u^{0} = \frac{1}{\sqrt{3}}\, [\Xi_{3/2}^{0}   + \sqrt{2} 
\Xi_{1/2}^{0} ]  \, , \quad \Xi_d^{0} = \frac{1}{\sqrt{3}} 
\,[\Xi_{1/2}^{0}  - \sqrt{2} \Xi_{3/2}^{0} ] \, . \label{qvsIstates} 
\eea 
These equations  can be immediately inverted to give 
\bea &&\Xi_{3/2}^{-}  = 
\frac{1}{\sqrt{3}}\, [\Xi_d^{-}  + \sqrt{2} \Xi_u^{-} ] \, , \quad 
\Xi_{1/2}^{-}  = 
\frac{1}{\sqrt{3}}\,[\Xi_u^{-}  - \sqrt{2}\Xi_d^{-}] \,  , \nonumber \\ 
&& \Xi_{3/2}^{0}  = \frac{1}{\sqrt{3}} \,[\Xi_u^{0}  - \sqrt{2} 
\Xi_d^{0}] \, , \quad \Xi_{1/2}^{0}  = \frac{1}{\sqrt{3}} \,[\Xi_d^{0}  + 
\sqrt{2}\Xi_u^{0}]  \, . \label{Ivsqstates} \eea 
 
\section{The mass matrices} 
\label{Sec3} 
We denote  the $2\times2$  mass matrices in the two charge 
sectors by ${\cal{M}}^{-}_q$, ${\cal{M}}^{-}_I$, 
${\cal{M}}^{0}_q$, ${\cal{M}}^{0}_I$ depending on the basis used, 
with the convention that the first row and column will denote 
$I=1/2$ in the isospin basis and $\Xi_d^{-} $ or $\Xi_u^{0}$ in the 
quark basis. If $q\bar{q}$ annihilation diagrams (i.e. violations 
of the OZI rule) as well as quark masses and electromagnetic (EM) 
interactions are neglected, all six states are exactly degenerate. 
Effects of OZI violations, quark masses and EM interactions will 
be added to first order. We will therefore discuss, in turn, OZI 
violations  neglecting quark mass and EM contributions, and then 
the isospin violations due to the latter two effects in the OZI 
limit. This approximation will be justified, a posteriori, if both kinds 
of corrections are small and comparable. 
 
\subsection{OZI violating contribution} 
\label{Subsec31} 
 
\begin{figure}[htb] 
\includegraphics[width=0.7\linewidth]{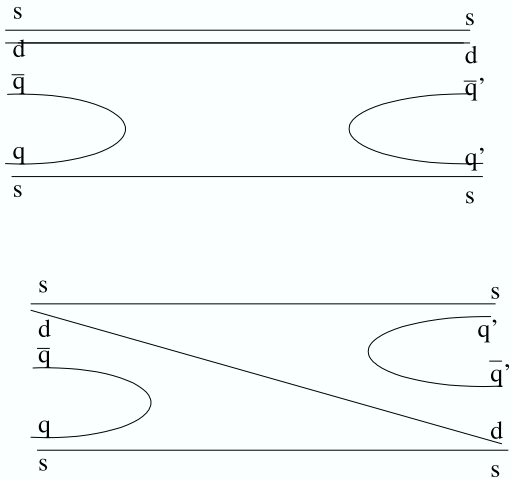}
\null 
\vspace{-1.cm} 
\caption{Diagrams that violate the OZI rule and mix different flavours 
in the $Q=-1$ sector.} 
\label{fig:FIGURE1} 
\end{figure}

The OZI-violating diagrams are shown in Fig.~\ref{fig:FIGURE1}. Clearly they 
contribute only to $I =1/2$ states. Their contribution to the mass 
matrix is thus particularly simple in the isospin basis in which, in 
both charge sectors, it reads: 
\begin{equation} 
\delta {\cal{M}}_I = \pmatrix{\delta & 0 \cr 0 & 0 \cr} \, , 
\label{IOZI} 
\end{equation} 
where, by definition, $\delta$ is the 
mass splitting between $I=1/2$ and $I =3/2$ states induced by OZI 
violations. We will comment later on the 
magnitude of $\delta$. In the quark basis (\ref{IOZI}) gives 
\begin{equation} 
\delta {\cal{M}}_q^- =  \pmatrix{\frac{2}{3} \delta & - 
\frac{\sqrt{2}}{3} \delta \cr 
 - \frac{\sqrt{2}}{3} \delta & \frac{1}{3} \delta \cr}  \, , 
\quad\delta {\cal{M}}_q^0 = \pmatrix{\frac{2}{3} \delta & 
\frac{\sqrt{2}}{3} \delta \cr 
   \frac{\sqrt{2}}{3} \delta & \frac{1}{3} \delta \cr}\, . 
\label{qOZI} 
\end{equation} 
OZI violations tend to align eigenstates along pure $SU(2)$ or $SU(3)$ 
representations. In the case of $SU(3)$, the explicit breaking due to the 
strange-quark mass supposedly induces strong $SU(3)$ mixing so that the 
true eigenstates are expected to be close to those with definite 
strange-quark content. This is the case for the ${\bf 8}+{\bf \bar{10}}$ 
 classification  of pentaquarks in the proposal of ref.~\cite{JW}. 
Here we are addressing, instead, the question of the relative magnitude of 
$\delta$ and of the explicit $SU(2)$ violating contributions 
that we are now going to discuss. 
 
\subsection{Isospin violations from quark masses and EM interactions} 
\label{Subsec32} 
 
Quark masses contribute to the QCD Hamiltonian with an iso-singlet 
piece proportional to $m = (m_u + m_d)/2 $ and an iso-triplet term 
proportional to $\delta m \equiv  (m_d - m_u)$. The former provides a common 
mass-term for all six states: being interested only in mass differences 
we will neglect it. The latter piece reads: 
\begin{equation} 
\delta H_m = -\frac{1}{2} \delta m~ \bar{q}\tau_3 q \, , 
\label{Hm} 
\end{equation} 
and transforms as an $I=1, I_3 =0$ operator. Its 
contributions to the mass matrices can be expressed in terms of 
three independent reduced matrix elements. 
 Two are the diagonal matrix elements corresponding to the $I=1/2$ 
and $I=3/2$ states and the third is 
the off-diagonal (transition) matrix element. Using standard CG 
coefficients, this gives: 
\begin{equation}  \delta{\cal{M}}_I^- = \pmatrix{\Delta^{(m)}_{1/2} & 
\Delta^{(m)}_{off} \cr \Delta^{(m)}_{off} & \Delta^{(m)}_{3/2} 
\cr} \, ,\quad \delta{\cal{M}}_I^0  =   \pmatrix{- 
\Delta^{(m)}_{1/2} & \Delta^{(m)}_{off} \cr \Delta^{(m)}_{off} & - 
\Delta^{(m)}_{3/2} \cr} \, . \label{mIMM} \end{equation} 
 
We now  impose the constraint 
 that comes from insisting that the OZI rule holds true, namely we require 
that, in the quark basis and in each charge sector, 
there is no off-diagonal (i.e.\ quark-flavour mixing) 
contribution. It is easy to check that such a constraint requires 
a single relation between the diagonal and off-diagonal entries in 
(\ref{mIMM}), namely:
\begin{equation} 
\Delta^{(m)}_{off}  = \sqrt{2} (\Delta^{(m)}_{3/2} - 
\Delta^{(m)}_{1/2})\, . \label{off} 
\end{equation} 
Using such a relation, the $\delta{\cal{M}}_I$ contributions to the 
mass matrices in the quark basis read: 
\begin{eqnarray}  \delta{\cal{M}}_q^- = \pmatrix{2 \Delta^{(m)}_{1/2} - 
\Delta^{(m)}_{3/2} & 0 \cr 0 & 2 \Delta^{(m)}_{3/2} - 
\Delta^{(m)}_{1/2} \cr} \, , \label{mqMMM}\\\nonumber\\ 
\delta{\cal{M}}_q^0 =\pmatrix{\Delta^{(m)}_{3/2} - 2 \Delta^{(m)}_{1/2}& 0\cr 
0 & \Delta^{(m)}_{1/2} - 2 \Delta^{(m)}_{3/2}  \cr}  \, . 
\label{mqMMZ} 
\end{eqnarray} 
The above general result can be compared with the one obtained by 
simply counting the number of $u$ and $d$ quarks in each hadron. 
The resulting mass matrices are indeed of the 
form~(\ref{mqMMM})--(\ref{mqMMZ}) with the additional constraint 
$\Delta^{(m)}_{1/2} = 5 \Delta^{(m)}_{3/2}$. 
 
Let us turn now to EM effects. Since the electromagnetic 
current is a mixture of $I=0$ and $I =1$ terms, a virtual EM 
contribution  will involve, to order $\alpha$, $I=0,1,2$ terms in 
the effective Hamiltonian.  The $I=1$ contribution 
can be treated exactly as the $\delta{\cal{M}}_I^{-/0}$ 
contributions we have just discussed. It is enough to make the 
following replacements in the mass-matrices~(\ref{mqMMM}) and~(\ref{mqMMZ}): 
\begin{equation} 
\Delta^{(m)}_{1/2}  \rightarrow \Delta_{1/2} \equiv 
\Delta^{(m)}_{1/2}  +  \Delta^{(em)}_{1/2} \, ,\qquad 
\Delta^{(m)}_{3/2}  \rightarrow \Delta_{3/2} \equiv 
\Delta^{(m)}_{3/2} +  \Delta^{(em)}_{3/2}\, . \label{I1repl} 
\end{equation} 
By contrast, the $I=0$ piece will in general give different contributions 
to the $I=1/2$ and $I=3/2$ diagonal matrix elements. We shall denote 
by $\Delta_0$ the difference between the former and the latter. Finally, 
the $I=2$ piece of the EM Hamiltonian can only contribute to the diagonal 
$I=3/2$ entry, $\Delta_2$, and to the off-diagonal ones, $\Delta_{2\,off}$. 
Asking as before the validity of the OZI rule allows to express the 
transition matrix element in terms of $\Delta_0$ and $\Delta_2$ through 
the relation $\Delta_{2\,off}=\sqrt{2}~(\Delta_2 -\Delta_0)$. 
 
\subsection{Adding up all the contributions} 
 
We can now collect all the non-trivial contributions to the mass 
matrices (to be added to a common matrix ${\cal{M}}_0$ proportional to the 
unit matrix) and write their final form as follows: 
\bea 
\hspace{-.5cm}\delta{\cal{M}}_I^- &=&\pmatrix{\delta+\Delta_{1/2}+ 
\frac{1}{2}\Delta_0& \sqrt{2} 
(\Delta_{3/2} - \Delta_{1/2} + \Delta_2 - \Delta_0)\cr \sqrt{2} (\Delta_{3/2} 
- \Delta_{1/2} + \Delta_2 -   \Delta_0) & \Delta_{3/2} 
+ \Delta_2 -\frac{1}{2} \Delta_0 \cr} \, , 
\label{fMM1}\\\nonumber \\ \vspace*{5mm}\hspace{-.5cm}\delta{\cal{M}}_I^0 &=& 
\pmatrix{\delta - \Delta_{1/2} + \frac{1}{2} \Delta_0& \sqrt{2}(\Delta_{3/2} - 
\Delta_{1/2} - \Delta_2 + \Delta_0)\cr \sqrt{2} (\Delta_{3/2} - \Delta_{1/2} 
- \Delta_2 + \Delta_0) & - \Delta_{3/2} + \Delta_2 - 
\frac{1}{2} \Delta_0 \cr} \, , \label{fMM2}\\\nonumber \\ 
\vspace*{5mm} \hspace{-.5cm}\delta{\cal{M}}_q^-&=&\pmatrix{\frac{2}{3} \delta 
+ 2 \Delta_{1/2} - \Delta_{3/2} - \Delta_2 
+ \frac{3}{2} \Delta_0 & - \frac{\sqrt{2}}{3} 
\delta \cr - \frac{\sqrt{2}}{3} \delta & \frac{1}{3} \delta + 2 
\Delta_{3/2} - \Delta_{1/2} +2 \Delta_2 -\frac{3}{2} \Delta_0 \cr}, 
\label{fMM3}\\\nonumber \\ 
\vspace*{5mm}\hspace{-.5cm}\delta{\cal{M}}_q^0 &=&  \pmatrix{\frac{2}{3}\delta 
- 2 \Delta_{1/2} + \Delta_{3/2} - \Delta_2 + \frac{3}{2} \Delta_0 & 
\frac{\sqrt{2}}{3} 
\delta \cr \frac{\sqrt{2}}{3} \delta & \frac{1}{3} \delta - 2 
\Delta_{3/2} + \Delta_{1/2} +2 \Delta_2 -  \frac{3}{2} \Delta_0 \cr}. 
\label{fMM4}\eea 
One can check that simple toy models of Coulomb energy effects (like the one 
where quarks are taken to be equidistant~\cite{LW}) give  $\Delta_2 = 
\Delta_0$. However this is not a necessity. For instance, in a model 
where the average distance of quark pairs is different than the 
one between the quarks and the antiquark, it is possible 
to generate a non vanishing $\Delta_2 - \Delta_0$. 
 
Note that these mass matrices provide immediately, in each charge sector, an 
average mass $\bar{M}^-$, $\bar{M}^0$: 
\begin{equation} 
\bar{M}^- = M_0 + \frac{1}{2}(\delta +\Delta_{1/2} + \Delta_{3/2} 
+ \Delta_2) \, , \quad\bar{M}^0 = M_0 + \frac{1}{2}(\delta 
-\Delta_{1/2} - \Delta_{3/2}  + \Delta_2) \, , ~ \label{AVM} 
\end{equation} 
giving a  gap 
$\Delta_{1/2} + \Delta_{3/2}$ between the average 
mass in the two doublets. On the other hand, splitting and mixing 
within each sector only depends on $\delta$ and on the 
combinations 
\begin{equation} 
\delta^{-} \equiv \Delta_{1/2} - \Delta_{3/2} - 
\Delta_2 + \Delta_0 \, , \qquad 
\delta^{0} \equiv -\Delta_{1/2}+ \Delta_{3/2} - \Delta_2 + \Delta_0 
\label{DMZ} 
\end{equation} 
for the $Q=-1$ and  $Q=0$ sectors, respectively. 
 
We also give, within our first-order approximation, the mass of the 
two pure $I=3/2$ states. They receive contributions that are completely 
fixed by group theory in terms of $\Delta_0$, $\Delta_{3/2}$ and $ \Delta_2$. 
One finds: 
\begin{equation} 
M^{--} = M_0 + 3\Delta_{3/2}-\Delta_2-\frac{1}{2}\Delta_0\, ,\quad M^{+}=M_0 
- 3 \Delta_{3/2} - \Delta_2 - \frac{1}{2} \Delta_0 \, , 
\label{EXM} 
\end{equation} 
As a check note that, for $\Delta_{1/2} = 5 \Delta_{3/2}$ one gets identical 
quark-mass contributions to the mass of states having the same number of 
$u+\bar{u}$ and $d+\bar{d}$ quarks (such as $\Xi^{--}$ and $\Xi^{0}_d$), 
in agreement with the naive quark-counting approximation. 
 
One should also note the following relations: 
\bea 
&&({\bar M}^- + \bar{M}^0) - (M^{--} +  M^{+}) = \delta + 3  \Delta_2 
+ \Delta_0 \,, \label{DMB} \\ && 
M^{--} -  M^{+} = 6  \Delta_{3/2} \, , \label{DME} \\ && 
3({\bar M}^- - \bar{M}^0) - ( M^{--} - M^{+}) = 3(\Delta_{3/2} - 
\Delta_{1/2}) \,. 
\label{DMS} 
\eea 
 
Summarizing, we have been able to express all  mass matrix elements 
of the six $\Xi$ states in terms of a common mass $M_0$ and of {\it five} 
parameters $\delta,\Delta_0, \Delta_{1/2},\Delta_{3/2},\Delta_2$. 
Since, in principle, one can measure six masses and two mixing 
angles,  our scheme makes {\it two} testable predictions. 
 
\section{Eigenstates and their phenomenology} 
\label{Sec4} 
 
In order to determine the mass eigenstates and eigenvalues (hence 
the mixing angles) in the two charge sectors it is useful to 
rewrite the mass-matrices in the isospin basis in the form: 
\bea 
{\cal M}_I^- &=& \bar{M}^- \times 
1\!\!1+\pmatrix{(\delta+\delta^{-})/2 & -\sqrt{2}\, \delta^{-} \cr - 
\sqrt{2} \, \delta^{-} & - (\delta + \delta^{-})/2 } \, , 
\label{fIMM1} \\\nonumber \\{\cal M}_I^0 &=& \bar{M}^0  \times 
1\!\!1 +  \pmatrix{(\delta + \delta^{0})/2 & \sqrt{2}\,\delta^{0} \cr 
\sqrt{2}\, \delta^{0} & - (\delta + \delta^{0})/2 } 
\, . \label{fIMM2} \eea 
Diagonalization is of course trivial. The mass eigenvalues are given by: 
\bea M^-_{1,2} &=& \bar{M}^- \pm 
\frac{1}{2} \sqrt{(\delta + \delta^{-})^2 + 
8 (\delta^{-})^2}  \,  , \label{Meigen1}\\ \nonumber \\ 
M^0_{1,2} &=&   \bar{M}^0 \pm \frac{1}{2} \sqrt{(\delta + \delta^{0})^2 + 
8 (\delta^{0})^2}  \, , \label{Meigen2} 
\eea 
while, defining the 
mixing angle as $\Xi_1={\rm cos}\,\theta ~ \Xi_{1/2}+{\rm 
sin}\,\theta ~ \Xi_{3/2} $, we find: 
\bea {\rm tan}\, \theta^- &=& 
\frac{1}{2\sqrt{2} \epsilon^-} \left[ (1+\epsilon^-) - \sqrt{(1 
+2\epsilon^- + 9 (\epsilon^-)^2}\,\right]\, ,\label{mixingangles1}\\ 
{\rm tan}\, \theta^0  &=&   - \frac{1}{2\sqrt{2} \epsilon^0} 
\left[(1+\epsilon^0)-\sqrt{(1+2 \epsilon^0+9(\epsilon^0)^2}\,\right] \, , 
\label{mixingangles2} \eea 
where we have set: 
\begin{equation} 
\epsilon^-  = \frac{\delta^{-}}{\delta}  \, , \quad \epsilon^0 
= \frac{\delta^{0}}{\delta} \, . \label{epsilon} 
\end{equation} 
It is easy to check that the 
mixing angles approach the expected values in the two limiting 
cases: zero mixing when $ \epsilon=0$ and maximal (or $SU(2)$ ``ideal") 
mixing (meaning here ${\rm tan}\, \theta^- = -1/\sqrt{2}$ and 
${\rm tan}\,\theta^0=-\sqrt{2} $) for $\epsilon^- \rightarrow \infty$ and 
$\epsilon^0 \rightarrow -\infty$, respectively. 
The mixing angles depend only on the two quantities $\epsilon^- $ 
and $ \epsilon^0$. The latter can be determined, for instance,  from the 
two independent ratios that one can form from  the differences 
$M_2^- -M_1^-$, $M_2^0 -M_1^0$ and the combination~(\ref{DMS}). 
 
Although we are still far, experimentally, from being able to estimate 
the mixing angles this way, we claim that we can nevertheless already 
exclude very small mixing angles  (pure isospin states). The argument 
goes as follows. Assume mixing to be small in both sectors, 
i.e.\ $\epsilon \ll 1$. In this case the observed mass splitting between 
the $\Xi^{--}$ and the $\Xi^{-}$ (which, being observed to decay in 
$\Xi^{0*}(1530) \pi^-$, is identified~\cite{JW} with the $\Xi^{-}_{1/2}$ 
 state  of eq.~(\ref{Ivsqstates})) 
implies $\delta = - 7 \pm 3$ MeV. This, however, is also the 
expected order of magnitude of the isospin breaking parameters 
$\delta^-$ and $\delta^0$. Indeed, since $3 \delta^-$ is the 
isospin-breaking  contribution to the difference between the two 
diagonal matrix elements of the matrices~(\ref{fMM3}) 
and~(\ref{fMM4}), a  naive  quark-counting argument, together with 
a simple model of Coulomb interaction energy  with equidistant 
quarks and antiquark~\cite{LW,RV2}, leads to the estimate 
\begin{equation} 
\delta^- = -\delta^0=\frac{2}{3}\,\delta m +\frac{1}{9} 
\alpha \left<\frac{1}{r}\right>\, . 
\label{EXPDM} 
\end{equation} 
A rough estimate of the r.h.s. of eq.~(\ref{EXPDM}), using~\cite{LW} 
$\delta m \sim 4.5~{\rm MeV}$~\footnote{Lattice estimates give 
$\delta m(2{\rm{Gev}})_{\overline{MS}}=2.3\pm .2\,({\rm{stat}}) 
\pm .5\,({\rm{syst}})~{\rm{Mev}}$~\cite{LAT}, where most of the large 
systematic error comes from unquenching. We remark that the number in 
eq.~(\ref{EXPDM}) is the down-up mass difference at a substantially 
smaller scale, of order $ \Lambda_{\rm{QCD}}$.} and 
$\left<\frac{1}{r}\right> \sim 240~{\rm MeV}$, gives 
\begin{equation} 
\delta^- = -\delta^0\sim 3.2~{\rm MeV} \sim - 0.46~ \delta \, , 
\label{EST} 
\end{equation} 
with a large error. In any case~(\ref{EST}) is in contradiction with 
the assumption of very small $|\epsilon|$'s. A slightly more 
sophisticated model, where the average distance of quarks 
inside a diquark is smaller than the average diquark-diquark or 
diquark-antiquark distance, relates $\delta^- = -\delta^0$ to 
$m(n) - m(p) + m(\Xi^-) - m(\Xi^0) \sim 8\, {\rm{MeV}}$, giving 
$\delta^- = -\delta^0 \sim - 0.4~ \delta$. 
 
\begin{figure}[htb] 
\vspace{-0.4cm} 
\includegraphics[width=1.0\linewidth]{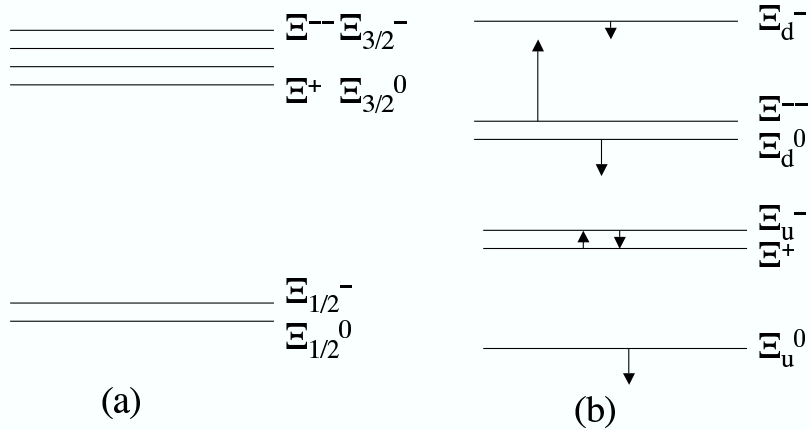} 
\vspace{-3.cm}
\caption{Schematic structure of the six energy levels in the two extreme 
cases: (a) very small isospin mixing,  i.e.\ $\epsilon^0,\epsilon^-<<1$; 
(b) almost ideal mixing. In (b) the line levels take into account quark 
and diquark mass differences, while the arrows show direction and 
relative magnitude of Coulomb effects for each state in the 
diquark picture. The relative normalization of the two 
isospin-breaking effects is not fixed} \label{fig:FIGURE2}\end{figure}

Obviously, once mixing is non-negligible, the above estimates of 
$\delta^- $ and $\delta^0$ have to be reconsidered. Within a definite 
model for isospin breaking effects, a determination of $\delta$ is possible 
from the measured splitting between the $\Xi^{--}$ and the observed 
$\Xi^{-}$ (now interpreted as one of the two negative-charge mass 
eigenstates). 
This eventually leads to an evaluation of $\epsilon^-$ and the mixing 
angle $\theta^-$ using (\ref{mixingangles1}). We find that, in some cases, 
$\delta$ comes out so small that almost ideal mixing is expected. However, 
more precise data on the mass splittings and a better theory of isospin 
breaking terms is needed before one can make definite predictions. 
In order to show how sensitive the mass splittings are to mixing we show, 
in Fig.~\ref{fig:FIGURE2}, the structure of the six mass eigenstates in the 
two extreme situations of very small and very large (i.e. ideal) mixing. 
Eqs.~(\ref{mixingangles1}) and~(\ref{mixingangles2}) imply that 
for $|\epsilon | > 0.3$, mixing angles are sufficiently large for 
their effects to be easily detectable. Indeed, for not too small 
values of $\theta^-$ ($\theta^0$), we should see two 
quasi-degenerate peaks in the $Q=-1$ ($Q=0$) sector with large 
violation of isospin present in their decays. For instance, in the 
$Q=-1$ sector only one of the two peaks (the one which is 
prevalently the $\Xi_u^-=(ds)(us)\bar{u}$ quark state) should be 
easily visible in the $\Xi^{*0}\pi^-$ channel, while also the 
second peak (corresponding to a state close to the 
$\Xi_d^-=(ds)(ds)\bar{d}$ quark state) should appear in 
isospin-related channels with a $\pi^0$ in the final state. 
 
We conclude by mentioning that, in the pentaquark picture assumed here, 
isospin mixing is also be expected to occur in the $S=-1$ sector.
In fact, since the states that can mix must differ by the replacement of a 
$u\bar u$ with a $d\bar d$ pair, the only other interesting doublet 
consists of the two neutral $\Sigma/\Lambda$-like $(ud)(qs)\bar q$ states. 
These are indeed the two lightest states with $Q=0$, $S=-1$ once 
$SU(3)$ mixing has lifted up those with three strange quarks. Similar 
phenomena should also occur in pentaquarks with one heavy quark (but not 
in those with a heavy antiquark). If the values of the $\epsilon$-like 
parameters we have introduced, will turn out to be in the appropriate 
range in some of the channels, rather spectacular phenomena will be able 
to put many of the current theoretical models to an interesting test. 
 
\vskip .2cm {\bf Acknowledgments} - We would like to thank L. Trentadue 
for his collaboration in the early stages of this work, Bob Jaffe for an 
interesting discussion on exotics in the Jaffe-Wilczek framework, and 
L. Giusti for correspondence on quark mass lattice data. One of us 
(G.C.R.) would like to thank the Humboldt Foundation for financial support.

\end{document}